\documentclass{PoS}

\title{
\vspace*{-2.5cm}
\begin{minipage}{\textwidth}
{\normalfont\small IFIC/15-72, LPN15-028
\hspace{\fill} October 2015
}\\
\end{minipage}\\[60pt]
  Higher-order QCD corrections to triple collinear splitting functions}

\ShortTitle{Higher-order QCD corrections to triple collinear splitting functions}

\author{\speaker{Germ\'an F. R. Sborlini}$^{a,b}$, Daniel de Florian$^a$ and Germ\'an Rodrigo$^b$\\\\
        $^a$Departamento de F\'\i sica and IFIBA, FCEyN, Universidad de Buenos Aires, 
(1428) Pabell\'on 1 Ciudad Universitaria, Capital Federal, Argentina.\\\
        $^b$Instituto de F\'{\i}sica Corpuscular, Universitat de Val\`{e}ncia -- 
Consejo Superior de Investigaciones Cient\'{\i}ficas, Parc Cient\'{\i}fic, E-46980 Paterna, Valencia, Spain.\\\\
        E-mail: \email{gfsborlini@df.uba.ar, deflo@df.uba.ar, german.rodrigo@csic.es}}

\abstract{We present splitting functions in the triple collinear limit at next-to-leading order in the strong coupling. We performed the computation in the context of massless QCD+QED, and
consider first collinear processes which include at least one photon. The IR divergent structure
of the multi-partonic splitting functions agrees with the Catani's formula. Consistency checks
based on symmetry arguments have been implemented and results for different configurations
have been cross-checked. Studying photon-started processes, we obtained very compact
results: this allowed us to simplify the expressions for the remaining splitting functions.}

\FullConference{The European Physical Society Conference on High Energy Physics\\
		22--29 July 2015\\
		Vienna, Austria}

\renewcommand{\thefigure}{\arabic{figure}}

\def\beq{\begin{equation}} \def\eeq{\end{equation}}
\def\beqn{\begin{eqnarray}} \def\eeqn{\end{eqnarray}}
\def\bom#1{{\mbox{\boldmath $#1$}}} \def\to{\rightarrow}
\def\nn{\nonumber}

\def\li#1{\mathrm{Li_2}\left(#1\right)}
\def\ln#1{\mathrm{log}\left(#1\right)}
\def\Eq#1{Eq.~(\ref{#1})}

\newcommand\DST{D_{\rm ST}}

\newcommand\IC{\textit{I}}

\renewcommand{\thefigure}{\arabic{figure}}

\def\ep{\epsilon}

\def\wp{\widetilde P}

\def\beq{\begin{equation}}
\def\eeq{\end{equation}}
\def\beeq{\begin{eqnarray}}
\def\eeeq{\end{eqnarray}}

\def\bom#1{{\mbox{$#1$}}}
\def\to{\rightarrow}

\newcommand{\la}{\langle}
\newcommand{\ra}{\rangle}

\def\nn{\nonumber}

\def\ID{1 \kern -.45 em 1}

\def\sp{{\bom {Sp}}}
\def\ket#1{|{#1}\ra}

\def\cbet0{b_0}

\begin{document}
\section{Introduction}
\label{sec:Introduction}
A proper understanding of the singular behaviour of scattering amplitudes in perturbative QFT is crucial to reach more accurate predictions. In particular, in the context of collider physics and QCD, infrared (IR) limits constitute an important part in the development of computational methods for higher-orders. In this presentation, we center the discussion in the collinear limit of scattering amplitudes and the calculation of \emph{splitting functions}.

Based on collinear factorization theorems \cite{Berends:1987me, Mangano:1990by}, when two or more external particles become almost parallel scattering amplitudes are expressed in terms of reduced amplitudes (which involve less particles) and universal factors, called \emph{splitting functions}. Some restrictions to this picture may be applied in special kinematical configurations. In fact, as shown in Refs. \cite{Catani:2011st,Forshaw:2012bi}, strict-collinear factorization~\cite{Collins:1989gx} guarantees the universality of splitting functions and their independence of the non-collinear particles only for time-like (TL) kinematics. In space-like (SL) kinematics, some correlations among collinear and non-collinear particles due to non-Abelian interactions might survive, thus breaking the traditional factorization concept.

Splitting functions are defined independently of the number of collinear particles and the perturbative order under consideration. These objects were first introduced for the double collinear limit in Ref. \cite{Altarelli:1977zs}. Since then, double-collinear splittings have been computed at one-loop and two-loop level, both for amplitudes and squared matrix-elements. In the last case, they are usually called \emph{Altarelli-Parisi} (AP) or \emph{splitting kernels}, since they control the perturbative evolution of parton distribution functions (PDFs) through the DGLAP equations. Moreover, higher-order corrections to AP kernels lead to a major improvement in the predictions obtained with parton shower generators.

The main topic of this presentation relies in the multiple collinear limit at higher-orders. These objects are essential ingredients of ${\rm N}^k$LO hadronic computations, with $k \geq 2$. Besides that, they play an interesting role in the perturbative generation of flavour asymmetries in PDFs \cite{Catani:2004nc, Rodrigo:2004kb}. Tree-level multiple collinear splittings were computed by many groups \cite{Campbell:1997hg, Catani:1998nv, DelDuca:1999ha, Birthwright:2005ak, Birthwright:2005vi, Catani:1999ss}. At one-loop level, there were only some partial results for $q \to q \bar{Q} Q$ \cite{Catani:2003vu}. Recently, the complete list of triple-collinear one-loop splitting functions for processes involving at least one photon was computed \cite{Sborlini:2014mpa, Sborlini:2014kla, Sborlini:2014hva}. Also, amplitude-level results for triple-collinear QCD processes at one-loop became available lately \cite{Badger:2015cxa}.

This work aims to briefly describe the computation of splitting functions. We set up the notation and some preliminary definitions in Sec. \ref{sec:properties}. After that, we describe the computation of (un)polarized splitting kernels in Sec. \ref{sec:results}, and we present some simplified results for photon-initiated processes. The conclusions and outlook are given in Sec. \ref{sec:conclusions}.

\section{Collinear limit: general properties}
\label{sec:properties}
Let's start with a generic $n$-particle process with $m$ almost collinear particles. We define the sets $C=\{1,2, \ldots, m\}$ and $NC=\{m+1,\ldots, n\}$ of indices associated with collinear and non-collinear particles, respectively. Momenta are labelled as $p_i$, with $p_i^2=0$ because we consider massless partons. Also, it is useful to introduce the scalar-products $s_{i j} = 2 \, p_i \cdot p_j$ and $s_{i,j} = \left(p_i +p_{i+1} + \ldots + p_{j} \right)^2 = p_{i,j}^2$. As stated in Ref. \cite{Catani:2011st}, strict-collinear factorization is guaranteed to be valid only in TL kinematics, i.e. when $s_{ij} \geq 0$ for every $i,j \in C$. 

The description of kinematical variables is very relevant to properly approach to the collinear limit. For this reason, we introduce an arbitrary light-like vector $n^{\mu}$ and define
\beq
\wp^{\mu} = p_{1,m}^{\mu} - \frac{s_{1,m}}{2 \; n \cdot \wp} \, n^{\mu} \, , \quad \quad  z_i = \frac{n \cdotp p_i}{n\cdot \wp}~, \qquad i \in C~,
\label{ptilde}
\eeq
that corresponds to the collinear direction and the longitudinal momentum fractions, respectively. Notice that $\wp^{\mu}$ is a light-like vector which exactly agrees with $p_{1,m}^{\mu}$ in the collinear configuration. On the other hand, momentum conservation in splitting process implies that $\sum_{i\in C} z_i = 1$.

In order to obtain manifest collinear factorization formulae, it is requested to work in a physical gauge, like the \textit{light-cone gauge} (LCG) \cite{Pritchard:1978ts}. This gauge choice avoids the introduction of non-physical degrees of freedom, which allows to express on-shell internal lines in terms of external states thus \textit{splitting} the original process. In consequence, in the limit $s_{1,m} \to 0$, the leading part of the scattering amplitude is given by
\beqn
\ket{{\cal A}\left(p_1, \ldots, p_n\right)} &\simeq& \sp_{a \to a_1 \ldots a_m}(p_1 ,\ldots, p_m; \wp) \, \otimes \, \ket{{\cal A}(\wp, p_{m+1}, \ldots, p_n)}~,  
\label{FACTORIZACIONLOmultiple1}
\eeqn
where $\sp_{a \to a_1 \ldots a_m}$ is the splitting amplitude and the symbol $\otimes$ denotes a sum over colors and polarizations of the intermediate state, i.e. the \textit{parent parton}. When the parent-parton is a vector-like particle, spin correlations become relevant. Thus, it is important to keep this information in order to have a complete description of the collinear limit. For this reason, we define the polarized vector splitting functions, which are obtained from the tensor product of two amputated splitting matrices, i.e.
\beqn
P^{\mu \nu}_{V \to a_1 \ldots a_m} &\equiv& \left( \frac{s_{1,m}}{2 \;\mu^{2\ep}} \right)^{m-1} \; \left(\sp^{\mu}_{V \to a_1 \ldots a_m}\right)^{\dagger} \sp^{\nu}_{V \to a_1 \ldots a_m} + \, {\rm h.c. }\, ,
\label{PpolarizedNLOdefinition}
\eeqn
where there is an implicit sum over the colors and spins of the external partons, and we average over the parent parton's colors. To recover the unpolarized splitting, we just contract with the polarization tensor in the LCG, $d_{\mu \nu}(\wp,n)$, and divide by the number of polarizations,
\beqn
\la \hat{P}_{V \to a_1 \cdots a_m} \ra &=& \frac{1}{\omega} \, d_{\mu \nu}(\wp,n) \,  P^{\mu \nu}_{V \to a_1 \ldots a_m} \, \, ,
\label{EcuacionDESPOLARIZA} 
\eeqn
where $\omega=2(1-\epsilon)$ since we are working in $D=4-2\epsilon$ space-time dimensions \cite{Bollini:1972ui,'tHooft:1972fi}. Of course, \Eq{PpolarizedNLOdefinition} can be extended for fermion-started processes, although there are some subtleties related with the presence of helicity-violating terms in DREG schemes\footnote{A detailed discussion about the physical interpretation of helicity-violating interactions in DREG is available in Refs. \cite{Sborlini:2014kla,Sborlini:2013jba}. For the sake of simplicity, we just mention that fermion-started polarized splitting functions are proportional to $\delta_{s \, s'}$ times the unpolarized splitting function.}.

To conclude this section, let's discuss briefly the $\epsilon$-pole structure of higher-order splitting functions. At one-loop level, $\sp_{a \to a_1 \ldots a_m}^{(1)}$ is decomposed as
\beqn
\sp^{(1)}_{a \to a_1 \dots a_m}(p_1, \dots, p_m; \wp) &=& \sp^{(1)\,{\rm fin.}}_{a \to a_1 \dots a_m} + \IC^{(1)}_{a \to a_1 \dots a_m}(p_1,\dots,p_m;\wp) \, \sp^{(0)}_{a \to a_1 \dots a_m}~,
\label{SeparaSP1}
\eeqn
where $\sp^{(1)\,{\rm fin.}}_{a \to a_1 \dots a_m}$ contains only finite terms. The insertion operator $\IC^{(1)}_{a \to a_1 \dots a_m}$ is described by Catani's formula \cite{Catani:2003vu,Catani:1996vz} and fixes the $\epsilon$-pole structure of the splitting amplitude. It is important to notice that \Eq{SeparaSP1} becomes slightly different for SL-kinematics because the divergent part could include correlations among collinear and non-collinear partons \cite{Catani:2011st, Forshaw:2012bi}. For this reason, we restrict our computations to the TL region.

\section{Triple collinear splittings at NLO}
\label{sec:results}
In order to compute triple collinear splitting functions at one-loop level in QCD$+$QED, we follow the techniques described in Refs. \cite{Sborlini:2014mpa, Sborlini:2014kla}. Let's summarize the procedure. The first step consists in calculating the amputated scattering amplitude for the process $a \to a_1 a_2 a_3$, considering off-shell kinematics for the parent parton (i.e. $p_{1,3}^2=s_{123}>0$). Then, we build a basis of rank-$2$ tensors using $\{p_i^{\mu},(\eta^{D})^{\mu \nu},n^{\mu}\}$ and we project the scattering amplitudes in this basis: this leads to a system of linear equations. Due to gauge symmetry and momentum conservation, the final result can be expressed using the reduced basis
\beqn
\nn f_{1}^{\mu\nu} &=& \eta_{\DST}^{\mu\nu} \, , \quad\quad f_{2}^{\mu\nu} = \frac{p_1^{\mu}p_2^{\nu}+p_1^{\nu}p_2^{\mu}}{s_{123}} \, ,  
\\ f_{3}^{\mu\nu} &=& 2 \frac{p_1^{\mu}p_1^{\nu}}{s_{123}} \, , \quad\quad f_{4}^{\mu\nu} = \left(f_{3}^{\mu\nu}\right)_{1 \leftrightarrow 2} \, , \quad\quad f_{5}^{\mu\nu} = \frac{p_1^{\mu}p_2^{\nu}-p_1^{\nu}p_2^{\mu}}{s_{123}} \, .
\eeqn
Once we obtain the coefficient for each element of the reduced tensorial basis, we perform an $\epsilon$ expansion up to ${\cal O}(\epsilon^0)$ and subtract the poles according to the decomposition proposed in \Eq{SeparaSP1}. Thus, the polarized splitting function becomes
\beqn
P^{(1)\,{\rm fin.},\mu \nu}_{a \to a_1 a_2 a_3} &=& c^{a \to a_1 a_2 a_3} \left[ \sum_{j=1}^{4} \, A^{(1)\,{\rm fin.}}_j f_{j}^{\mu \nu} \, + \, A^{(1)\,{\rm fin.}}_5 f_{5}^{\mu \nu} \right] \, ,
\label{EquacionDescomposicion}
\eeqn
where $c^{a \to a_1 a_2 a_3}$ is a global normalization factor. Finally, we perform a simplification in the coefficients $A^{(1)\,{\rm fin.}}_i$ exploiting the classification of functions by their transcendental weight. According to this idea, rational functions have weight $0$, whilst $\ln{x}^n$, ${\rm Li}_n(x)$, $\pi^n$ or $\zeta_n$ have transcendental weight $n$. In QCD, the finite part of one-loop computations in the limit $\epsilon \to 0$ involve up to weight $2$ functions. So, we rewrite the coefficients in \Eq{EquacionDescomposicion} as
\beqn
\nn A^{(1)\,{\rm fin.}}_j &=& \sum_{i=0}^2 {\cal C}_j^{(i)}  \, + (1 \leftrightarrow 2) \,    \ {\rm for} \ j\in\left\{1,2\right\} \, , 
\\ A^{(1)\,{\rm fin.}}_3 &=& \sum_{i=0}^2 {\cal C}_3^{(i)}  \, , \quad\quad A^{(1)\,{\rm fin.}}_4 = \left(A^{(1)\,{\rm fin.}}_3\right)_{1 \leftrightarrow 2} \, , \quad\quad A^{(1)\,{\rm fin.}}_5 = \sum_{i=0}^2 {\cal C}_5^{(i)}  \, - (1 \leftrightarrow 2)  \, ,
\eeqn
where ${\cal C}_j^{(i)}$ includes only functions of transcendental weight $i$.

\subsection{$\gamma \to q \bar{q} \gamma$}
\label{ssec:photon}
This is the simplest possible process in the triple-collinear limit. Using the previous notation and the dimensionless variables $x_i=s_{jk}/s_{123}$, the rational terms are given by
\beqn
{\cal C}_{1}^{(0)} &=& \frac{1-x_1}{x_1} \left(\frac{8 (1-x_1)}{x_2} + 1 \right), \quad\quad {\cal C}_{2}^{(0)} = \frac{4}{1-x_3} \left( \frac{1-x_1}{x_1\, x_2} - 1 \right) - \frac{2}{1-x_1}, 
\\ {\cal C}_{3}^{(0)} &=& \frac{1}{x_1\, x_2} \left(\frac{4 (1-x_2+x_2^2)}{1-x_3} + \frac{(1-x_2)^2}{1-x_1} + 15 - x_2 \right), \ \ {\cal C}_{5}^{(0)} = - \frac{2}{x_1} \left( \frac{1}{1-x_1} -\frac{2}{1-x_3} \right), \ \ \ \
\eeqn
whilst those involving transcendental functions are
\beqn
{\cal C}_{1}^{(1)} &=& \frac{1-x_2}{x_2} \left( \frac{2x_3-x_2}{1-x_1} \log (x_1) + \frac{2 x_3}{1-x_3}  \log (x_3) \right)~, 
\\ \nn {\cal C}_{2}^{(1)} &=& \frac{2}{x_1\, x_2} \Bigg[ \frac{1}{1-x_1} \left( \frac{2x_3}{x_2}-\frac{2x_1\, x_2 + x_3}{1-x_1}\right) \log (x_1) 
\\ &+&\frac{2}{1 - x_3} \left( \frac{x_3}{x_1} + \frac{x_3 (1 - x_1) - x_1 x_2}{1 - x_3} \right) \log (x_3) \Bigg]~,
\\ \nn {\cal C}_{3}^{(1)} &=& \frac{(1 - x_2)^2}{x_1\, x_2 (1 - x_1)} \left( \frac{1}{1 - x_1} + \frac{2}{x_2} \right) \log(x_1) + \frac{2 x_3 - x_1}{x_1^2\, x_2} \log(x_2) 
\\ &+& \frac{2}{(1 - x_3)^2} \left(\frac{2 (2 - x_2) x_3}{x_1\, x_2} + \frac{x_3^2}{x_1^2} + \frac{1}{x_2^2} -  2 \right) \log(x_3)~, 
\\ {\cal C}_{5}^{(1)} &=&  \frac{2}{1-x_1} \left( \frac{2x_3}{x_1\, x_2} - \frac{1}{1-x_1} \right) \log (x_1) + \frac{4}{x_1 (1-x_3)^2}  \log (x_3)~,
\\ {\cal C}_{1}^{(2)} &=& \frac{2}{x_1\, x_2} \left[(1-x_3) x_3 \left(1-\frac{1}{x_2}\right)-(1-x_1)^2\right] \, {\cal R}\left(x_1,x_3\right)~, 
\\ {\cal C}_{2}^{(2)} &=& \frac{4}{x_2^3}\left(1-\frac{(1-x_2)^2}{x_1}\right) \, {\cal R}\left(x_1,x_3\right)~, 
\\ {\cal C}_{3}^{(2)} &=& - \frac{2}{x_1 x_2} \left[\left(2+\frac{(1-x_2)^2}{x_2^2}\right) {\cal R}\left(x_1,x_3\right) + \left(1+\frac{x_3^2}{x_1^2}\right) {\cal R}\left(x_2,x_3\right) \right]~, 
\\ {\cal C}_{5}^{(2)} &=& -\frac{4 x_3}{x_1 \, x_2^2} \, {\cal R}\left(x_1,x_3\right)~,
\eeqn
where ${\cal R}(x_i,x_j)$ is a combination of weight $2$ transcendental functions, given by
\beqn
{{\cal R}\left(x_i,x_j\right)} &=& \frac{\pi^2}{6}-\ln{x_i}\ln{x_j}-\li{1-x_i}-\li{1-x_j}  \, .
\label{DefinicionCALR}
\eeqn
It is related to the scalar-box integral and we notice that ${\cal R}(x_i,1-x_i)=0$ due to Euler's reflection formula: this situation takes place when one sub-energy goes to zero faster than $s_{1,3}$. Besides that, all these expressions are independent of $z_i$. This behaviour is related with the general fact that color singlets are invariant under ${\rm SU}(3)_{\rm C}$ transformations. In other words, photon-started splitting processes can be computed using the covariant gauge, thus avoiding to introduce the quantization vector $n^{\mu}$ inside loop-integrals. Of course, LO splitting functions for this kind of processes might still depend on $z_i$ due to the kinematics of the collinear configuration.

\section{Conclusions and outlook}
\label{sec:conclusions}
The collinear behaviour of scattering amplitudes is discussed in this presentation. For that purpose, we used splitting functions and described their calculation in the triple collinear limit. We restricted the attention to processes involving at least one photon, focusing in the treatment of one-loop QCD corrections. Since we worked in the TL-region, strict collinear factorization properties are fulfilled and we used this fact to compare the divergent structure of our results with the predictions of Catani's formula. Besides this, we calculated both polarized \cite{Sborlini:2014kla} and unpolarized \cite{Sborlini:2014mpa} splittings using independent codes, which was crucial to implement consistency-check between both sets of expressions. 

On the other hand, we found useful to analyse the structure of photon-initiated splitting functions, because they involved very compact expressions. This led to a significant simplification of the remaining splitting functions, because the Abelian contributions to QCD processes are obtained through the replacement of gluons by photons. Moreover, this fact allowed to implement more cross-checks among our computations.

  \subsection*{Acknowledgments}
This work is partially supported by UBACYT, CONICET, ANPCyT, the
Research Executive Agency (REA) of the European Union under
the Grant Agreement number PITN-GA-2010-264564 (LHCPhenoNet),
by the Spanish Government and EU ERDF funds
 (grants FPA2011-23778, FPA2014-53631-C2-1-P and CSD 2007-00042
Consolider Ingenio CPAN) and by GV (PROMETEUII/2013/007).












\end{document}